\documentclass[aps,amsmath,amssymb,preprintnumbers,groupedaddress,twocolumn]{revtex4}
\usepackage{graphicx,psfrag,hyperref}
\usepackage{amsmath,bbm,array,amsfonts,graphicx,wrapfig,lscape,float,mathtools,multirow,longtable}
\usepackage[dvipsnames]{xcolor}
\usepackage{collcell}

\newcommand{\be}{\begin{equation}}
\newcommand{\ee}{\end{equation}}
\newcommand{\beq}{\begin{equation}}
\newcommand{\beql}[1]{\begin{equation}\label{#1}}
\newcommand{\eeq}{\end{equation}}
\newcommand{\ba}{\begin{array}}
\newcommand{\ea}{\end{array}}
\newcommand{\bea}{\begin{eqnarray}}
\newcommand{\beal}[1]{\begin{eqnarray}\label{#1}}
\newcommand{\eea}{\end{eqnarray}}
\newcommand{\ben}{\begin{enumerate}}
\newcommand{\een}{\end{enumerate}}
\newcommand{\bean}{\begin{eqnarray*}}
\newcommand{\eean}{\end{eqnarray*}}
\newcommand{\eref}[1]{(\ref{#1})}
\newcommand{\sref}[1]{\S\ref{#1}}
\newcommand{\tref}[1]{Table~\ref{#1}}
\newcommand{\nn}{\nonumber}

\newcommand{\fref}[1]{Figure \ref{#1}}
\newcommand{\btab}[1]{\begin{tabular}{#1}}
\newcommand{\etab}{\end{tabular}}

\newcommand{\comment}[1]{}

\newcommand{\ud}{\mathrm{d}}

\newcommand{\qed}{\nobreak \ifvmode \relax \else
      \ifdim\lastskip<1.5em \hskip-\lastskip
      \hskip1.5em plus0em minus0.5em \fi \nobreak
      \vrule height0.75em width0.5em depth0.25em\fi}

\begin{document}

\preprint{UNIST-MTH-24-RS-04, RIKEN-iTHEMS-Report-24}

\title{Combinatorial and Algebraic Mutations of Toric Fano 3-folds \\ and Mass Deformations of $2d$ $(0,2)$ Quiver Gauge Theories}

\author{Dongwook Ghim${}^{a}$} 
\author{Minsung Kho${}^{b}$}
\author{Rak-Kyeong Seong${}^{b,c}$}
\email{dongwook.ghim@riken.jp, minsung@unist.ac.kr, seong@unist.ac.kr}

\affiliation{\it ${}^{a}$
Interdisciplinary Theoretical and Mathematical Sciences Program (iTHEMS), RIKEN, \\
2-1 Hirosawa, Wako, Saitama 351-0198, Japan \\
${}^{b}$ 
Department of Mathematical Sciences, and \\ 
 ${}^{c}$ 
Department of Physics,\\ 
Ulsan National Institute of Science and Technology,\\
50 UNIST-gil, Ulsan 44919, South Korea
}

\begin{abstract}
We argue that algebraic and combinatorial polytope mutations of Fano 3-folds can be identified with mass deformations of associated $2d$ $(0,2)$ supersymmetric gauge theories realized by brane brick models. 
These are Type IIA brane configurations that realize a large family of $2d$ worldvolume theories on probe D1-branes at toric Calabi-Yau 4-folds. 
We show that brane brick models that are related by mass deformations associated to algebraic and combinatorial polytope mutations of Fano 3-folds have mesonic moduli spaces with the same number of generators. 
We show that mesonic flavor charges of these generators form convex reflexive lattice polytopes that are dual to the toric diagrams of the Fano 3-folds. 
The generating function of mesonic gauge invariant operators, also known as the Hilbert series of the mesonic moduli space, appears to be identical for such brane brick models under a particular refinement 
originating from the $U(1)_R$ charges in the brane brick model following the mass deformation.
\end{abstract} 
\maketitle
\noindent

\section{Introduction}

The worldvolume theories of D1-branes probing a toric Calabi-Yau 4-fold form a large family of $2d$ $(0,2)$ supersymmetric gauge theories realized by brane brick models \cite{Franco:2015tna,Franco:2015tya,Franco:2016nwv,Franco:2016qxh}.
These IIA brane configurations consist of D4-branes suspended between a NS5-brane wrapping a holomorphic surface $\Sigma$ on a 3-torus $T^3$, given by
\beal{newton-poly}
\Sigma \, : \,  P(x,y,z) = 0 \,,
\eea
where $P(x,y,z)$ is the Newton polynomial in $x,y,z\in \mathbb{C}^*$ of the toric diagram $\Delta$ associated to the probed toric Calabi-Yau 4-fold. 

\begin{table}[h!]
\centering
\begin{tabular}{|c|cc|cccccc|cc|}
\hline
\; & 0 & 1 & 2 & 3 & 4 & 5 & 6 & 7 & 8 & 9 \\
\hline 
D4 & $\times$ & $\times$ & $\cdot$ & $\times$ & $\cdot$ & $\times$ & $\cdot$ & $\times$ & $\cdot$ & $\cdot$
\\
NS5 & $\times$ & $\times$ & \multicolumn{6}{c|}{------ $\Sigma$ ------} & $\cdot$ & $\cdot$
\\
\hline
\end{tabular}
\caption{Type IIA brane configuration for brane brick models.}
\label{t_brane}
\end{table}

\begin{figure*}
  \includegraphics[width=0.85\textwidth]{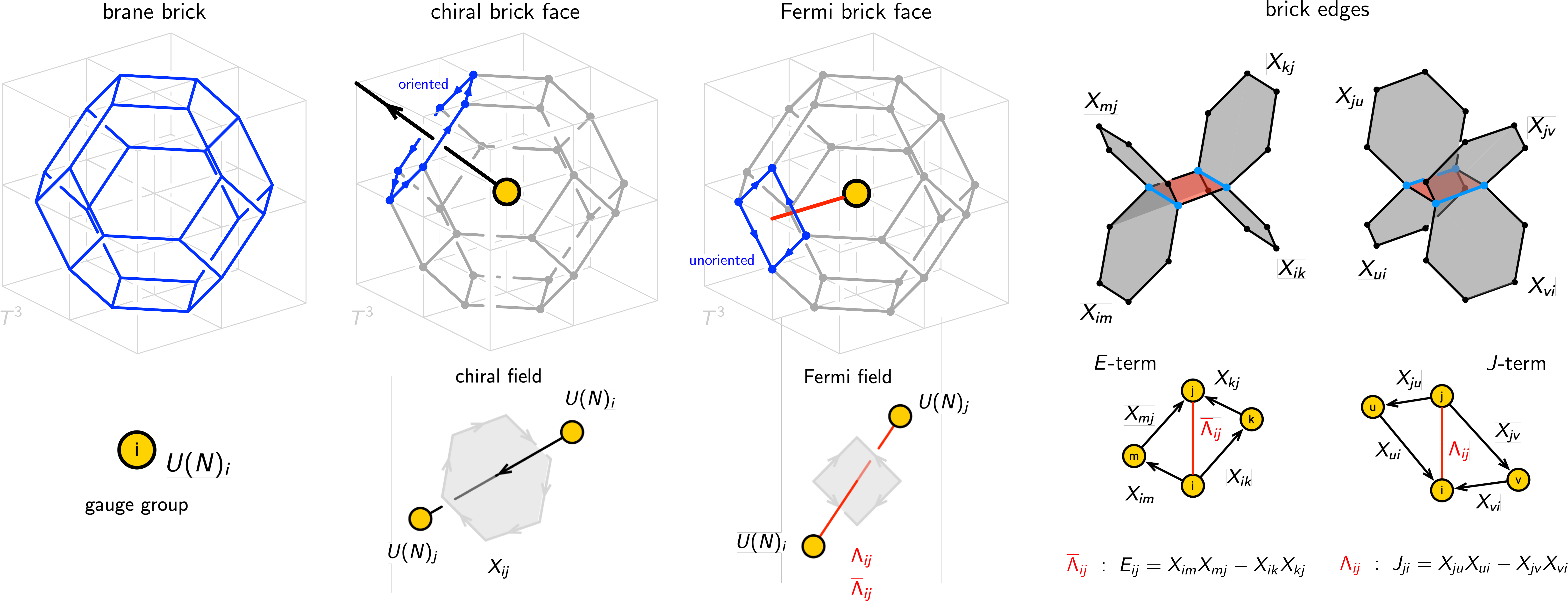}
  \caption{Dictionary between a brane brick model and its corresponding $2d$ $(0,2)$ supersymmetric gauge theory.\label{fig_01}}
\end{figure*}

The NS5-brane wrapping $\Sigma$ forms a tessellation of the 3-torus $T^3$ consisting of 3-dimensional polytopes known as \textit{brane bricks}, which have a boundary made of \textit{brick faces} and \textit{brick edges}.
These respectively can be identified with $U(N)$ gauge groups, chiral $X_{ij}$ and Fermi $\Lambda_{ij}$ fields, as well as $J$- and $E$-terms of the corresponding $2d$ $(0,2)$ theories following a dictionary first introduced in \cite{Franco:2015tna,Franco:2015tya} and illustrated in \fref{fig_01}.

In the following work, we are interested in a particular family of brane brick models that correspond to toric Calabi-Yau 4-folds associated to Fano 3-folds \cite{Franco:2022gvl}.
The toric diagram $\Delta$ of these toric Calabi-Yau 4-folds is a reflexive polytope in $\mathbb{Z}^3$, with a unique internal point at the origin $(0,0,0)$ and a reflexive dual $\Delta^\circ$ which is also a convex lattice polytope in $\mathbb{Z}^3$ \cite{batyrev1993dual,borisov1993towards,batyrev1999classification,batyrev1997dual,batyrev1996calabi}.
As classified by Kreuzer and Skarke \cite{kreuzer1997classification,kreuzer1998classification}, there are 4319 distinct reflexive polytopes in $\mathbb{Z}^3$ up to $Gl(3,\mathbb{Z})$ transformation on the vertices of $\Delta$.

The work in \cite{akhtar2012minkowski} introduced two seemingly unrelated but equivalent operations on reflexive toric diagrams of toric Calabi-Yau 4-folds, that relate two Fano 3-folds to each other. 
These operations are known as \textit{algebraic polytope mutations} and \textit{combinatorial polytope mutations} and it has been shown in \cite{akhtar2012minkowski} that amongst the 4319 reflexive polytopes, only 3025 exhibit these mutations and can be related to each other by such mutations. 

In this work, we argue that these mutations on reflexive lattice polytopes associated to toric diagrams of Fano 3-folds can be interpreted as a mass deformation of the corresponding $2d$ $(0,2)$ supersymmetric gauge theories realized by brane brick models. 
By explicitly computing the generating function of gauge invariant operators of the $2d$ $(0,2)$ theories, also known as the Hilbert series \cite{Benvenuti:2006qr,Hanany:2006uc,Butti:2007jv,Feng:2007ur,Hanany:2007zz} of the mesonic moduli space $\mathcal{M}^{mes}$ of the brane brick models, we also show that the number of gauge-invariant generators of $\mathcal{M}^{mes}$ is identical between the $2d$ $(0,2)$ theories related by the mass deformation and polytope mutations. 
By an appropriate unrefinement of the Hilbert series, we show that the Hilbert series are identical between the two brane brick models, which is also an observation made in \cite{akhtar2012minkowski} in terms of Ehrhart polynomials associated to the related toric diagrams. 

In the following work, we study properties of algebraic and combinatorial polytope mutations of toric Fano 3-folds in relation to mass deformation of brane brick models. 
We first review algebraic and combinatorial polytope mutations in section \sref{sec2} as first introduced in \cite{akhtar2012minkowski}, and summarize mass deformation for brane brick models in section \sref{s_mass} as studied in \cite{Franco:2023tyf}.
We then study in section \sref{s_comp} an explicit example of how the algebraic and combinatorial polytope mutations of the abelian orbifold of the form $\mathbb{C}^4/\mathbb{Z}_6$ with orbifold action $(1,1,2,2)$ \cite{Davey:2010px,Hanany:2010ne,Franco:2015tna} correspond to mass deformation between associated brane brick models. By doing so we make several observations of mathematical and physical properties on the correspondence between algebraic and combinatorial polytope mutations of toric Fano 3-folds and mass deformations of brane brick models, which we summarize in section \sref{s_conclusion}. 
\\

\section{Algebraic and Combinatorial Polytope Mutations}\label{sec2}

\paragraph{Reflexive Polytopes.}
Brane brick models are associated to toric Calabi-Yau 4-folds $\mathcal{X}$.
We can think of the toric Calabi-Yau 4-fold $\mathcal{X}$ as the complex cone over
a \textit{toric variety} $X(\Delta)$, which is given by a convex lattice polytope in $\mathbb{Z}^3$.
This polytope is what we call the \textit{toric diagram} $\Delta$ \cite{fulton1993introduction}.
Certain toric diagrams in $\mathbb{Z}^3$ are called \textit{reflexive}, if they have a \textit{dual polytope} $\Delta^\circ$, which is also a convex lattice polytope in $\mathbb{Z}^3$ \cite{batyrev1993dual,borisov1993towards,batyrev1999classification,batyrev1997dual,batyrev1996calabi}.
The dual polytope $\Delta^\circ$ of a convex lattice polytope $\Delta$ is defined as follows
\beal{e02a01}
\Delta^\circ
=
\{
\mathbf{v} \in \mathbb{Z}^3 ~|~ \mathbf{m} \cdot \mathbf{v} \geq -1 ~\forall \mathbf{m} \in \Delta
\}
~.~
\eea
We note that if the toric diagram $\Delta$ defining a toric variety $X(\Delta)$ is reflexive, the corresponding variety $X(\Delta)$ is \textit{Fano} \cite{nill2005gorenstein}. 
In this work, we concentrate on brane brick models associated with toric Fano 3-folds \cite{Franco:2022gvl}, whose toric diagrams are reflexive polytopes in $\mathbb{Z}^3$.
Based on the classification of Kreuzer and Skarke \cite{kreuzer1997classification,kreuzer1998classification}, we know that there are exactly 4319 distinct reflexive polytopes in $\mathbb{Z}^3$. There is a finite number of reflexive polytopes in each dimension $n$ up to a $Gl(n,\mathbb{Z})$ transformation. 
\\

\paragraph{Newton Polynomials and $Gl(3,\mathbb{Z})$ Transformations.}
Any two toric diagrams $\Delta$ in $n$ dimensions, whose vertices are mapped to each other by a $Gl(n,\mathbb{Z})$ transformation, are associated to the same toric variety $X(\Delta)$. 
Given a toric diagram $\Delta$ in a particular $Gl(n,\mathbb{Z})$ frame, 
we can write a Laurent polynomial of the following form to $\Delta$, 
\beal{es02a10}
P(x_1,\dots, x_n) = \sum_{\mathbf{v} \in \Delta}~ c_{\mathbf{v}}  ~\mathbf{x}^\mathbf{v} ~,~
\eea
where the vertices $\mathbf{v}=(v_1, \dots, v_n) \in \mathbb{Z}^n$, the fugacities $\mathbf{x}=(x_1, \dots, x_n)\in (\mathbb{C}^*)^n$, and $\mathbf{x}^\mathbf{v}= \prod_{i=1}^{n} x_i^{v_i}$.
We also have coefficients $c_{\mathbf{v}}\in \mathbb{C}^*$, which we interpret as a \textit{multiplicity} of the corresponding vertices $\mathbf{v}$.
These correspond to complex structure moduli of the mirror Calabi-Yau $(n+1)$-fold \cite{Hori:2000ck,Feng:2005gw,Franco:2016qxh}.
We refer to the Laurent polynomial in \eref{es02a10} as the \textit{Newton polynomial} $P$ of $\Delta$, if every $\mathbb{Z}^n$ lattice point in the convex hull of the toric diagram $\Delta$ has an associated term in $P$.

In terms of the Newton polynomial of a toric diagram $\Delta$, we can describe the $Gl(n,\mathbb{Z})$ transformation on $P$ of $\Delta$ as a transformation on the fugacities $x_1, \dots, x_n \in \mathbb{C}^*$.
In the case for $n=3$, we have
\beal{es02a10b}
&&
M_{ij} ~:~ P(x,y,z) \mapsto\nn\\
&&
 \hspace{0.2cm} P(x^{M_{11}} y^{M_{12}} z^{M_{13}} , x^{M_{21}} y^{M_{22}} z^{M_{23}}, x^{M_{31}} y^{M_{32}} z^{M_{33}}) ~,~
 \nn\\
\eea
where $M \in Gl(3,\mathbb{Z})$ with $i, j=1,2,3$. 
Here, we note that equivalence under a $Gl(n,\mathbb{Z})$ transformation on the Newton polynomial $P$ is stronger than 
under a $Gl(n,\mathbb{Z})$ transformation on the vertices of $\Delta$. 
This is because, the Newton polynomial $P$ of $\Delta$ keeps track of the multiplicities $c_{\mathbf{v}}$ of the vertices $\mathbf{v}$ of $\Delta$. 
\\

\paragraph{Algebraic Polytope Mutation.}

Let us consider the Newton polynomial $P(x,y,z)$ of a reflexive toric diagram $\Delta$, where we set the origin of $\Delta$ to be at $(0,0,0)$ and take the associated coefficient in $P(x,y,z)$ to be $c_{(0,0,0)}=0$. 
We assume that under a certain $Gl(3,\mathbb{Z})$ transformation of $P(x,y,z)$, 
the resulting Laurent polynomial takes the following general form,
\beal{es02a12}
P(x,y,z) = \sum_{m=a}^{b} C_m(x,y) z^m ~,~
\eea
where $a<0$ and $b>0$ and $C_m(x,y)$ with $a\leq m \leq b$ are sub-polynomials in $x,y$.

Based on Laurent polynomials of the form in \eref{es02a12},
 \cite{akhtar2012minkowski} introduced an \textit{algebraic polytope mutation} of $\Delta$ given by the following birational transformation $\varphi_A$ on $(x,y,z) \in (\mathbb{C}^*)^3$ as follows, 
 \beal{es02a15}
 \varphi_A
 ~:~
(x,y,z)&\mapsto& (x,y,A(x,y) z)
~,~
 \eea
where the Laurent polynomial $A(x,y)$ is chosen to be such that $A(x,y)^{-m}$ is a polynomial divisor of $C_m(x,y)$ in $P(x,y,z)$ for all $m$ in $a\leq m \leq -1$. 

 Under the above map, the resulting polynomial, which we denote as $\varphi_A P(x,y,z)$, takes the following form,
 \beal{es02a16}
\varphi_A P(x,y,z)= 
\sum_{m=a}^{b}
A(x,y)^{m}  ~C_m(x,y) ~
z^m
~.~
 \eea
When we interpret each term in the above Laurent polynomial $\varphi_A P(x,y,z) \equiv P^\vee(x,y,z)$ as a vertex corresponding to a new convex lattice polytope in $\mathbb{Z}^3$, including here again the origin $(0,0,0)$, we can interpret the resulting convex lattice polytope $\varphi_A \Delta \equiv \Delta^{\vee}$ as a toric diagram of a new deformed toric Calabi-Yau 4-fold. 

It was shown in \cite{akhtar2012minkowski,akhtar2016mutations} that if the deformed toric diagram $\Delta^\vee$ associated to the Laurent polynomial $P^\vee(x,y,z)$ in \eref{es02a16} is reflexive and corresponds to a toric Fano 3-fold, then the original toric diagram $\Delta$ has to be reflexive and associated to a toric Fano 3-fold as well. 
Interestingly, \cite{akhtar2012minkowski} showed also that among the 4319 reflexive polytopes in $\mathbb{Z}^3$ classified by Kreuzer and Skarke \cite{kreuzer1997classification,kreuzer1998classification}, only 3025 reflexive polytopes have a Laurent polynomial of the form in \eref{es02a12} that support an algebraic polytope mutation, whereas the other 1294 reflexive polytopes do not support algebraic polytope mutations. 
Such Laurent polynomials that support an algebraic polytope mutation are also referred to as \textit{Minkowski polynomials} in \cite{akhtar2012minkowski}.
 \\
 
\begin{figure*}
  \includegraphics[width=0.85\textwidth]{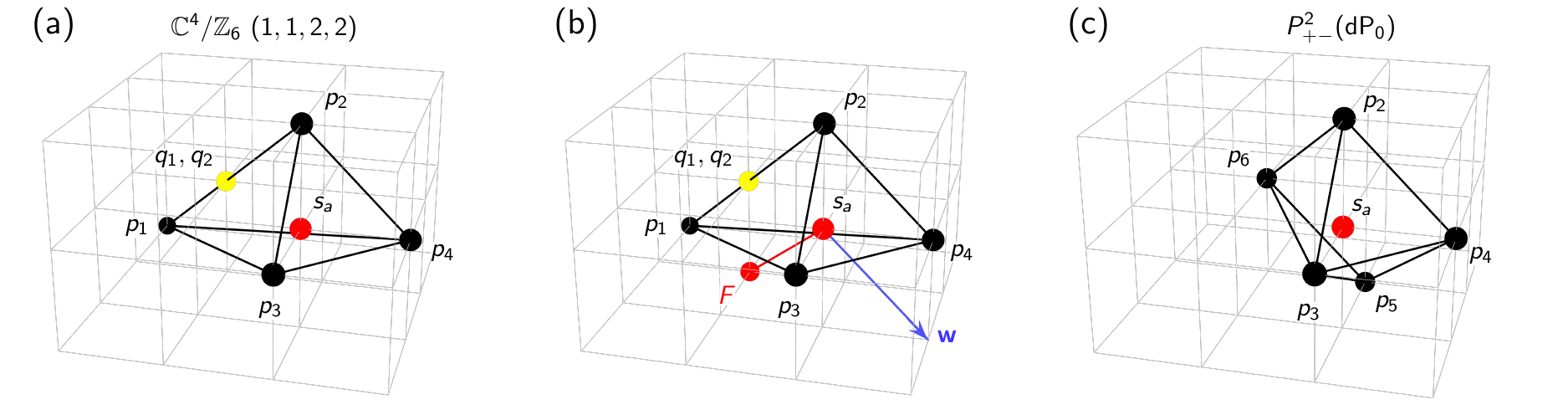}
  \caption{(a) The toric diagram for the $\mathbb{C}^4/\mathbb{Z}_6$ $(1,1,2,2)$ model before deformation, (b) the mutation vector $\mathbf{w}=(0,1,-1)$ and factor $F=\{(0,0,0), (-1,-1,-1)\}$, and (c) the toric diagram for the $P_{+-}^2(\text{dP}_0)$ model after deformation.\label{fig_02}}
\end{figure*}

 \noindent
 \textit{Example.}
 Let us consider the following Laurent polynomial for the abelian orbifold of the form $\mathbb{C}^4/\mathbb{Z}_6$ with orbifold action $(1,1,2,2)$ \cite{Davey:2010px,Hanany:2010ne,Franco:2015tna},
 \beal{es02a17}
 P(x,y,z) = x+ y+ z+ \frac{2}{x y} + \frac{1}{x^2 y^2 z} ~,~
 \eea
 where we have set the term corresponding to the internal point of the reflexive toric diagram to $0$.
 The corresponding toric diagram is shown in \fref{fig_02}(a). 
 Under the following $Gl(3,\mathbb{Z})$ transformation, 
 \beal{es02a18}
 M =
\left(
\begin{array}{@{}ccc@{}}
1 & 0 & 0 \\
0 & -1 & 1\\
0 & 0 & -1
 \end{array}
\right)~,~
 \eea
 the polynomial in \eref{es02a17} becomes
 \beal{es02a19}
 P(x,y,z) = 
 \frac{1}{y} z + x + \left( 1 + \frac{2y}{x} + \frac{y^2}{x^2} \right) \frac{1}{z}
 \eea
 which takes the general form of \eref{es02a12} with 
 \beal{es02a20}
 &
 C_{1} (x,y) = \frac{1}{y}
 ~,~
 C_{0} (x,y) = x 
 ~,~
 &
 \nn\\
 &
 C_{-1} (x,y) = 
 1 + \frac{2y}{x} + \frac{y^2}{x^2}
 ~,~
 &
 \eea
 and $a=-1, b=1$. 
 For the algebraic polytope mutation $\varphi_A$, we can choose the polynomial divisor 
$A(x,y)= 1 + \frac{y}{x}$, which transforms the Laurent polynomial in \eref{es02a19} into 
\beal{es02a20b}
\varphi_A P(x,y,z) = 
\left( \frac{1}{x} + \frac{1}{y} \right) z 
+ x 
+ \left( 1 + \frac{y}{x} \right) \frac{1}{z}
~,~
\nn\\
\eea
following \eref{es02a15}. 
Under the inverse $Gl(3,\mathbb{Z})$ transformation $M^{-1}$ based on \eref{es02a18}, we obtain the final polynomial of the form 
\beal{es02a20c}
\varphi_A P(x,y,z) = x+ y+z + \frac{1}{x y} + \frac{1}{x z} ~,~
\eea
whose corresponding toric diagram including the origin $(0,0,0)$ is shown in \fref{fig_02}(c).
This convex polytope is reflexive but not regular, and is part of the 4319 reflexive polytopes in $\mathbb{Z}^3$ \cite{kreuzer1997classification,kreuzer1998classification}. 
Following the naming convention in \cite{Franco:2022gvl}, we will refer to the corresponding toric Calabi-Yau 4-fold as $P_{+-}^2(\text{dP}_0)$. 
 \\
 
\paragraph{Combinatorial Polytope Mutation.}
Alternative to algebraic polytope mutation, we can directly mutate the toric diagram of a toric Calabi-Yau 4-fold using a combinatorial method introduced in \cite{akhtar2012minkowski}.
We call this method the \textit{combinatorial polytope mutation} $\mu_{\textbf{w}}$. 

First, let $\textbf{w}$ be a primitive lattice vector and $\Delta$ be a convex lattice polytope in $\mathbb{Z}^3$. 
Given $\textbf{w}$ and $\Delta$, we can define
\beal{es02a20}
h_{min} &=& \min\{\textbf{w} \cdot \textbf{v}~|~\textbf{v} \in \Delta \}~,~
\nn\\
h_{max} &=& \max\{\textbf{w}\cdot \textbf{v}~|~\textbf{v} \in \Delta \} ~,~
\eea
which measure the highest and lowest vertex in $\Delta$ along the direction set by $\textbf{w}$.
Then the $\textit{width}$ of the toric diagram $\Delta$ with respect to the vector $\textbf{w}$ is defined as,
\beal{es02a21}
\text{width}_\textbf{w} (\Delta) = h_{max}-h_{min}
\eea 
Accordingly, for each lattice point $\textbf{v} \in \mathbb{Z}^3$, we say that $\textbf{v}$ is at $\textit{height}$ $h$ if $\textbf{w}\cdot \textbf{v}=h$. 
For each height $h$ along vector $\textbf{w}$, we can define a hyperplane of lattice points in $\mathbb{Z}^3$ given by 
$\text{H}_{\textbf{w},h}=\{\textbf{x} \in \mathbb{Z}^3 ~|~ \textbf{w} \cdot \textbf{x}=h\}$. The set of vertices in $\Delta$ at height $h$ along vector $\textbf{w}$ is then defined as, 
\beal{es02a22}
{w}_{h} (\Delta) = \text{conv}(\text{H}_{\textbf{w},h} \cap \Delta) ~,~
\eea
where $\text{conv}$ refers to the convex hull of the vertices at height $h$.

Let us now assume that there is a set of vertices $F \in \mathbb{Z}^3$, also known as the \textit{factor}, with $\textbf{w} \cdot (F)=0$ such that there are polytopes $G_{h}$ at height $h$ satisfying the following relation,
\beal{es02a25}
H_{\textbf{w},h} \cap V(\Delta) \subseteq G_h+(-h) F \subseteq w_h (\Delta)
\eea
for every $h_{min} \leq h < 0$.
Here, $V(\Delta)$ refers to the extremal vertices of the toric diagram $\Delta$.
Furthermore, the sum in $G_h+(-h) F$ refers to the \textit{Minkowski sum}, which is defined as follows,
\beal{es01a02}
Q+R = \{\textbf{q}+\textbf{r}~| \textbf{q} \in Q~,~\textbf{r} \in R\}
~,~
\eea
where $Q$ and $R$ are two convex polytopes in $\mathbb{Z}^3$. 

Using the above definitions, we can introduce the $\textit{combinatorial mutation}$ $\mu_{\textbf{w}}(\Delta, F ;\{G_h\})$ of a toric diagram $\Delta$ as follows,
\beal{es02a26}
&&
\mu_{\textbf{w}}(\Delta,F;\{G_h\})
=
\nn\\
&& 
\hspace{0.5cm}
\text{conv} \Big(\bigcup_{h=h_{min}}^{-1} G_{h}~\cup~\bigcup_{h=0}^{h_{max}}(w_{h}(P)+F) \Big)~.~ \nn\\
\eea
Following \cite{akhtar2012minkowski}, the mutation leads to the same polytope independent of the choice of $G_{h}$. 
Accordingly, we can refer to the combinatorial mutation of $\Delta$ as $\mu_{\textbf{w}}(\Delta,F)$.  

As noted in \cite{akhtar2012minkowski}, for any algebraic mutation $\varphi_A$ of a reflexive toric diagram $\Delta$ as shown in \eref{es02a15} with corresponding Minkowski polynomial $P(x,y,z)$ of the form in \eref{es02a12}, there is always an associated combinatorial mutation $\mu_{\textbf{w}}(\Delta; F) = \text{Newt}(\varphi_A P(x,y,z))$ of $\Delta$ with $\textbf{w}=(0,0,1)$ and $F=\text{Newt}(A(x,y))$, where $h_{min}= a$ and $h_{max} = b$ in \eref{es02a20}. 
Here, $A(x,y)$ is the Laurent polynomial in the algebraic mutation in \eref{es02a15}, where $\text{Newt}(A(x,y))$ corresponds to the convex Newton polygon in $\mathbb{Z}^3$ associated to the Laurent polynomial $A(x,y)$.
The polytopes $G_h$ in the combinatorial mutation take the form,
\beal{es02a28}
G_h = \text{Newt} \left(\frac{C_m(x,y)}{A(x,y)^{-m}} \right) ~,~
\eea
where $h_{min} \leq m \leq -1$. 
\\

\noindent
\textit{Example.}
Let us now briefly discuss an example for a combinatorial polytope mutation and its corresponding algebraic polytope mutation. 
We take the example of the abelian orbifold of the form $\mathbb{C}^4/\mathbb{Z}_6$ $(1,1,2,2)$ whose Laurent polynomial is in \eref{es02a17}. The corresponding toric diagram $\Delta$ has the following lattice points in $\mathbb{Z}^3$, 
 \beal{es02a30}
\{
 (1,0,0), 
 (0,1,0),
 (0,0,1),
 (-1,-1,0),
 (-2,-2,-1)
 \}~,~
 \nn\\
 \eea
 where the single internal point of the reflexive polytope is taken to be $(0,0,0)$.
 By choosing, 
 \beal{es02a31}
 \textbf{w}= (0,1,-1),~
 F=\{
 (0,0,0),
 (-1,-1,-1)
 \},~
 \eea
 we can define the combinatorial mutation $\mu_{\textbf{w}}(\Delta, F)$ following \eref{es02a26}.
 We note here that along $\textbf{w}=(0,1,-1)$, the original toric diagram $\Delta$ has heights $-1 \leq h \leq 1$ with a polytope width of $\text{width}_{\textbf{w}}(\Delta) = 2$. 
 Accordingly, we only have a choice to make for the set of polytopes for $G_{-1}$, which need to satisfy the following condition from \eref{es02a25},
 \beal{es02a32}
 &&
\{ (0,0,1),(-2,-2,-1) \} ~\subseteq~ G_{-1}+F 
\nn\\
&&
\hspace{1cm}
~\subseteq~ \{ (0,0,1),(-2,-2,-1),(-1,-1,0) \} 
~,~
\nn\\
 \eea
 where we have,
 \beal{es02a33}
 &&
 H_{\textbf{w}, -1} \cap V(\Delta) = \{ (0,0,1),(-2,-2,-1) \}
 ~,~
 \nn\\
 &&
 w_{-1}(\Delta) = \{ (0,0,1),
 (-2,-2,-1),
 (-1,-1,0) \}
 ~.~
 \nn\\
 \eea
 Based on the above conditions, we fix
 \beal{es02a34}
 G_{-1} = \{
 (-1,-1,0),
 (0,0,1)
 \}~,~
 \eea
 such that the combinatorial mutation of $\Delta$ under $\textbf{w}$ and $F$ is given by,
 \beal{es02a35}
\mu_{\textbf{w}}(\Delta,F)
&=&
\text{conv}
\Big(G_{-1}~\cup w_{0} (P)~\cup (w_{1}(P)+F)\Big) 
\nn\\  
&=&
\text{conv}\Big(\{(1,0,0),(0,1,0),(0,0,1),
\nn\\ 
&& 
\hspace{1cm}  
(0,0,0),(-1,0,-1),(-1,-1,0)\}
\Big)
~.~
\nn\\
\eea
We can see here that the resulting convex lattice polytope $\mu_{\textbf{w}}(\Delta, F)$ as shown in \fref{fig_02}(c), has the corresponding Laurent polynomial $\varphi_A P(x,y,z)$ in \eref{es02a20c}, with the term associated to the internal point $(0,0,0)$ set to $0$. 
We therefore see that the toric diagram obtained from algebraic polytope mutation $\varphi_A \Delta \equiv \Delta^\vee$ is the same as the one obtained from combinatorial polytope mutation $\mu_{\textbf{w}}(\Delta,F)\equiv \Delta^\vee$, both corresponding to the toric diagram for $P^2_{+-}(\text{dP}_0)$ as shown in \fref{fig_02}(c).
\\

\section{Mass Deformations of Brane Brick Models}\label{s_mass}

\paragraph{Mesonic Moduli Space and Hilbert Series.}
Brane brick models representing $2d$ $(0,2)$ supersymmetric gauge theories have $J$- and $E$-terms of the following binomial form, 
\beal{es03a01}
&&
\Lambda_{ij} ~:~
J_{ji} = J_{ji}^{+} - J_{ji}^{-} ~,~
\nn\\
&&
\overline{\Lambda}_{ij} ~:~
E_{ij} = E_{ij}^{+} - J_{ji}^{-} ~,~
\eea
where $J_{ji}^\pm$ and $E_{ij}^\pm$ correspond to monomial products of chiral fields $X_{ij}$. 
The moduli space of gauge invariant operators subject to the $J$- and $E$-terms of the $2d$ $(0,2)$ theory, which is also known as the \textit{mesonic moduli space} $\mathcal{M}^{mes}$ \cite{Franco:2015tna,Franco:2015tya,Franco:2016nwv,Franco:2016qxh}, takes the following form for abelian brane brick models,
\beal{es03a02}
\mathcal{M}^{mes}
= 
\text{Spec}~
(\mathbb{C}[X_{ij}] / \mathcal{I}^{\text{Irr}}_{JE} ) // U(1)^{G-1}
\eea
where $\mathbb{C}[X_{ij}]$ is the coordinate ring formed by the chiral fields $X_{ij}$, 
$\mathcal{I}_{\text{Irr}}$ is the irreducible component of the toric ideal formed by the binomial $J$- and $E$-terms, 
and $U(1)^{G-1}$ are the independent $U(1)$ gauge groups of the $2d$ $(0,2)$ theory. 
Here, $G$ is the total number of $U(1)$ gauge groups, where one $U(1)$ decouples, and $i,j=1, \dots, G$ are the $U(1)$ gauge group labels.

By expressing the $n_\chi$ chiral fields in terms of GLSM fields $p_a$ \cite{Witten:1993yc}, which are identified combinatorially as brick matchings in the brane brick model \cite{Franco:2015tya}, the mesonic moduli space $\mathcal{M}^{mes}$ can be expressed as the following symplectic quotient, 
\beal{es03a03}
\mathcal{M}^{mes} = \text{Spec} ~(\mathbb{C}[p_a] // Q_{JE} ) // Q_{D} ~,~
\eea
where $a=1, \dots, c$ counts over the GLSM fields $p_a$ and $c$ is the total number of GLSM fields.
$Q_{JE}$ and $Q_D$ are the $U(1)$ charge matrices on the GLSM fields due to the $J$- and $E$-terms as well as the $D$-terms, respectively. 
Following the \textit{forward algorithm} for brane brick models \cite{Franco:2015tna,Franco:2015tya}, these charge matrices can be obtained as follows, 
\beal{es03a04}
(Q_{JE})_{(c-G-3) \times c} &=& \ker P_{n_\chi \times c} ,~
\nn\\
\overline{d}_{(G-1)\times n_\chi} &=& (Q_{D})_{(G-1)\times c} \cdot P^{t}_{c\times n_\chi} ,~
\eea
where $P$ is the $(n_\chi \times c)$-dimensional brick matching matrix encoding the map between GLSM fields and chiral fields of the brane brick model, and $\overline{d}$ is the reduced incidence matrix of the quiver for the brane brick model \cite{Franco:2015tna,Franco:2015tya}.
Overall, 
we can identify the mesonic moduli space $\mathcal{M}^{mes}$ of the abelian $2d$ $(0,2)$ theory as the toric Calabi-Yau 4-fold corresponding to the brane brick model. 
The vertices of the toric diagram $\Delta$ are encoded in the following coordinate matrix, 
\beal{es03a05}
(G_t)_{4\times c} = \ker (Q_t)_{(c-4) \times c}
\eea
where $Q_t= (Q_{JE}, Q_D)$.
We note here that each vertex in $\Delta$ is associated to at least one of the GLSM fields $p_a$ of the brane brick model. 

The \textit{Hilbert series} \cite{Benvenuti:2006qr,Hanany:2006uc,Butti:2007jv,Feng:2007ur,Hanany:2007zz} captures the algebro-geometric structure of the mesonic moduli space $\mathcal{M}^{mes}$ and allows us with the use of \textit{plethystics} to identify the generators and the defining relations of $\mathcal{M}^{mes}$. 
In terms of the symplectic quotient description of the mesonic moduli space $\mathcal{M}^{mes}$ in \eref{es03a03}, we can calculate the Hilbert series using the following Molien integral formula, 
\beal{es03a10}
&&
g(y_a, t; \mathcal{M}^{mes}) = 
\nn\\
&&
\hspace{1cm}
\prod_{i=1}^{c-4}
\oint_{|z_i| = 1} \frac{\ud z_i}{2\pi i z_i} 
\prod_{a=1}^{c}
\frac{1}{1-t_a \prod_{j=1}^{c-4} z_j^{(Q_t)_{ja}}}
~.~
\nn\\
\eea
The fugacities $t_a$ used in the Hilbert series count the degrees in the GLSM fields $p_a$.  
We can introduce a change of fugacities of the Hilbert series as follows, 
\beal{es03a11}
t_a = y_1^{q^1_a} y_2^{q^2_a} y_3^{q^3_a} t^{r_a} ~,~
\eea
where 
for each GLSM field $p_a$ we have the charges
$(q^1_a, q^2_a,q^3_a) \in \mathbb{R}^3$ under the mesonic flavor symmetry with corresponding fugacities $y_1,y_2, y_3$, and the $U(1)_R$ charge $r_a$ with corresponding fugacity $t$.
\\

\paragraph{Mass Deformations.}
Mass terms in $2d$ $(0,2)$ theories are gauge invariant quadratic terms in the Lagrangian involving a chiral-Fermi pair. 
As a result, in order for the theory to admit a mass deformation, the quiver $Q$ of the $2d$ $(0,2)$ theory needs to contain pairs of chiral and Fermi fields between the same pair of quiver gauge nodes.
Accordingly, the following shows the general form of possible mass terms one can add to the $J$- and $E$-terms of a brane brick model \cite{Franco:2023tyf}, 
\beal{es03a20}
&&
(\Lambda_{ij}, X_{ij}) \in Q ~:~ 
J_{ji}^\prime = J_{ji} ~,~ E_{ij}^\prime = \pm m X_{ij} + E_{ij} ~,~
\nn\\
&&
(\overline{\Lambda}_{ij}, X_{ji}) \in Q ~:~ 
J_{ji}^\prime = \pm m X_{ji} + J_{ji} ~,~ E_{ij}^\prime = E_{ij} ~,~
\nn\\
\eea
where $J_{ji}^\prime$ and $E_{ij}^\prime$ are the $J$- and $E$-terms after the mass deformation, and $J_{ji}$ and $E_{ij}$ are the $J$- and $E$-terms before the mass deformation. 
When the Fermi fields $\Lambda_{ij}$ are integrated out, the mass deformations in \eref{es03a20} lead to the following chiral field replacements, 
\beal{es03a21}
X_{ij} = \mp \frac{1}{m} (E_{ij}^+ - E_{ij}^- ) ~,~
X_{ji} = \mp \frac{1}{m} (J_{ji}^+ - J_{ji}^-) ~.~
\nn\\
\eea
In order to ensure that the resulting $J$- and $E$-terms after the mass deformation form a toric variety, we have to make sure that all of them are binomial relations. 
Based on the results in \cite{Franco:2023tyf}, this can be guaranteed by introducing mass deformations for pairs of chiral-Fermi pairs of the form $(\Lambda_{ij}, X_{ij})$ and $(\Lambda_{kl}, X_{kl})$, or $(\overline{\Lambda}_{ij}, X_{ji})$ and $(\overline{\Lambda}_{kl}, X_{lk})$, with the correct sign assignments on the respective mass terms.
As outlined in \cite{Franco:2023tyf}, additional re-definitions of Fermi-chiral interaction terms in the Lagrangian of the $2d$ $(0,2)$ theory ensure the toric condition of the $J$- and $E$-term after mass deformation. 
These re-definitions take the following general form, 
\beal{es03a22}
\Lambda_{ij}^\prime \cdot X_{jk}^\prime = \Lambda_{ij} \cdot (X_{jk} + \sum_h c_h^{(jk)} X_{jh} X_{hk} )~,~
\eea
where $X_{jk}^\prime$ and $\Lambda_{ij}^\prime$ indicate the new fields after the re-definition, and $c_h^{(jk)}$ are coefficients specific to the re-definition. 

In this work, we observe that mass deformation of brane brick models corresponding to toric Fano 3-folds with reflexive toric diagrams can be associated to algebraic and combinatorial polytope mutation of the toric Fano 3-folds. 
This is because under mass deformation certain vertices of the toric diagram $\Delta$ of the original toric Fano 3-fold change their relative position, giving rise to a new toric Fano 3-fold with a corresponding new toric diagram $\Delta^\vee$, which is the same obtained under algebraic and combinatorial polytope mutation. 
These vertices that change their relative position correspond to \textit{moving brick matchings} in the mass deformation as identified in \cite{Franco:2023tyf}.
Besides these brick matchings, we also have \textit{massive brick matchings} \cite{Franco:2023tyf}, which contain all chiral fields that become massive during the mass deformation. 
In the following section, we show that the algebraic and combinatorial polytope mutation connecting $\mathbb{C}^4/\mathbb{Z}_6$ $(1,1,2,2)$ with $P_{+-}^{2}(\text{dP}_0)$
has a corresponding mass deformation connecting the associated brane brick models.
\\

\begin{figure}
  \includegraphics[width=0.45\textwidth]{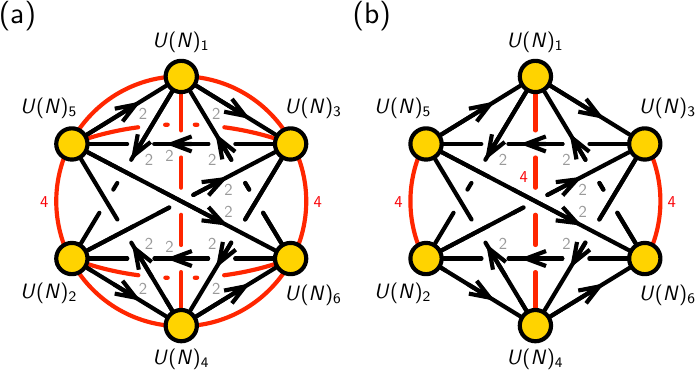}
  \caption{The quiver diagrams for (a) the $\mathbb{C}^{4}/\mathbb{Z}_6$ $(1,1,2,2)$ model before mutation, and (b) the $P_{+-}^{2}(\text{dP}_0)$ model after mutation. \label{fig_045}}
\end{figure}

\section{Algebraic and Combinatorial Mutations and Mass Deformations}\label{s_comp}

Let us consider the brane brick model associated to the abelian orbifold of $\mathbb{C}^{4}/\mathbb{Z}_6$ with orbifold action $(1,1,2,2)$ whose quiver is shown in \fref{fig_045}(a), and the $J$- and $E$-terms are given below, 
\beal{es05a10}
\begin{array}{rrcccc}
\Lambda^{1}_{14} :  & J_{41}^{1} & = & Y_{46} \cdot Z_{61} &- & Z_{45} \cdot Y_{51} \\  
\overline{\Lambda}^{1}_{14} :  & E_{14}^{1} & = &  D_{13} \cdot X_{34} &- & X_{12} \cdot D_{24} \\
\Lambda^{2}_{14} :   & J_{41}^{2} & = &  X_{45} \cdot Y_{51} & - & Y_{46} \cdot X_{61} \\
\overline{\Lambda}_{14}^{2} :   & E_{14}^{2} & = & D_{13} \cdot Z_{34} & - & Z_{12} \cdot D_{24} \\
\Lambda^{1}_{25} :  & J^{1}_{52} &=& Y_{51} \cdot Z_{12} &- & Z_{56} \cdot Y_{62} \\ 
\overline{\Lambda}^{1}_{25} :  & E^{1}_{25} &= & D_{24} \cdot X_{45} &-& X_{23} \cdot D_{35} \\ 
\Lambda^{2}_{25} :  & J^{2}_{52} &= & X_{56} \cdot Y_{62} & - & Y_{51} \cdot X_{12} \\ 
\overline{\Lambda}^{2}_{25} :  & E^{2}_{25} & = & D_{24} \cdot Z_{45} &-& Z_{23} \cdot D_{35} \\ 
\Lambda^{1}_{36} :  & J^{1}_{63} & = & Y_{62} \cdot Z_{23} &-& Z_{61} \cdot Y_{13} \\  
\overline{\Lambda}^{1}_{36} :  & E^{1}_{36} &= & D_{35} \cdot X_{56} &-& X_{34} \cdot D_{46} \\
\Lambda^{2}_{36} :  & J^{2}_{63} & = & X_{61} \cdot Y_{13} & - & Y_{62} \cdot X_{23} \\  
\overline{\Lambda}^{2}_{36} :  & E^{2}_{36} &=& D_{35} \cdot Z_{56} &-& Z_{34} \cdot D_{46} \\
\Lambda^{1}_{41} :  & J^{1}_{14} &=& Y_{13} \cdot Z_{34} &-& Z_{12} \cdot Y_{24} \\ 
\overline{\Lambda}^{1}_{41} :  & E^{1}_{41} & = & D_{46} \cdot X_{61} & - & X_{45} \cdot D_{51} \\
\Lambda^{2}_{41} : &  J^{2}_{14} & = & X_{12} \cdot Y_{24} &-& Y_{13} \cdot X_{34} \\ 
\overline{\Lambda}^{2}_{41} : & E^{2}_{41} & = & D_{46} \cdot Z_{61} &-& Z_{45} \cdot D_{51} \\
\Lambda^{1}_{52} :  & J^{1}_{25} & = &Y_{24} \cdot Z_{45} &-& Z_{23} \cdot Y_{35} \\  
\overline{\Lambda}^{1}_{52} :  & E^{1}_{52} & = & D_{51} \cdot X_{12} &-& X_{56} \cdot D_{62} \\
\Lambda^{2}_{52} :  & J_{25} &=&  X_{23} \cdot Y_{35} &-& Y_{24} \cdot X_{45} \\ 
\overline{\Lambda}^{2}_{52} :  & E_{52} &=& D_{51} \cdot Z_{12} &-& Z_{56} \cdot D_{62} \\
\Lambda^{1}_{63} :  & J^{1}_{36} &=&  Y_{35} \cdot Z_{56} &-& Z_{34} \cdot Y_{46} \\ 
\overline{\Lambda}^{1}_{63} :  & E^{1}_{63} & = & D_{62} \cdot X_{23} &-& X_{61} \cdot D_{13} \\
\Lambda^{2}_{63} :  & J^{2}_{36} & = & X_{34} \cdot Y_{46} &-& Y_{35} \cdot X_{56} \\  
\overline{\Lambda}^{2}_{63} :  & E^{2}_{63} & = &  D_{62} \cdot Z_{23} &-& Z_{61} \cdot D_{13} \\
\Lambda_{15} :  &  J_{51} &= & Z_{56} \cdot X_{61} & - & X_{56} \cdot Z_{61}  \\  
\overline{\Lambda}_{15} :  & E_{15} & = & D_{13} \cdot Y_{35} & - & Y_{13} \cdot D_{35} \\
\Lambda_{26} :  & J_{62} & = & Z_{61} \cdot X_{12} &- &X_{61} \cdot Z_{12} \\  
\overline{\Lambda}_{26} :  & E_{26} & = & D_{24} \cdot Y_{46} &-& Y_{24} \cdot D_{46} \\ 
\Lambda_{31} :  & J_{13} & = & Z_{12} \cdot X_{23} &-& X_{12} \cdot Z_{23} \\  
\overline{\Lambda}_{31} :  & E_{31} & = & D_{35} \cdot Y_{51} &-& Y_{35} \cdot D_{51} \\
\Lambda_{42} :  & J_{24} & = & Z_{23} \cdot X_{34} &-& X_{23} \cdot Z_{34} \\  
\overline{\Lambda}_{42} :  & E_{42} & = & D_{46} \cdot Y_{62} &-& Y_{46} \cdot D_{62} \\
\Lambda_{53} :  & J_{35} & = & Z_{34} \cdot X_{45} &-& X_{34} \cdot Z_{45} \\  
\overline{\Lambda}_{53} :  & E_{53} & = & D_{51} \cdot Y_{13} &-& Y_{51} \cdot D_{13} \\
\Lambda_{64} :  & J_{46} & = & Z_{45} \cdot X_{56} &-& X_{45} \cdot Z_{56}  \\ 
\overline{\Lambda}_{64} :  &  E_{64} & = & D_{62} \cdot Y_{24} &-& Y_{62} \cdot D_{24} \\
\end{array} 
~.~
\eea
Using the forward algorithm for brane brick models \cite{Franco:2015tna,Franco:2015tya}, we are able to calculate the brick matching matrix as follows, 
\beal{es05a11}
P=
\resizebox{0.32\textwidth}{!}{$
\left(
\ba{c|cccc|cc|cccccc|ccc}
\; &  p_1 & p_2 & p_3 & p_4 & q_1 & q_2 & s_1 & s_2 & s_3 & s_4 & s_5 & s_6 & o_1 & o_2 & o_3  \\
\hline
 D_{13} &  0 & 0 &  1 &  0 &  0 &  0 &  0 &  0 &  0 &  0 &  1 &  1 &  0 &  0 &  1 \\ 
 D_{24} &  0& 0 &  1 &  0  &  0 &  0 &  0 &  0 &  0 &  1 &  0 &  1 &  0 &  1 &  0 \\ 
 D_{35} &  0 & 0 &  1 &  0 &  0 &  0 &  0 &  0 &  1 &  1 &  0 &  0 &  1 &  0 &  0 \\ 
 D_{46} &  0 & 0 &  1 &  0 &  0 &  0 &  0 &  1 &  1 &  0 &  0 &  0 &  0 &  0 &  1 \\ 
 D_{51} &  0 & 0 &  1 &  0 &  0 &  0 &  1 &  1 &  0 &  0 &  0 &  0 &  0 &  1 &  0 \\ 
 D_{62} &  0 & 0 &  1 &  0 &  0 &  0 &  1 &  0 &  0 &  0 &  1 &  0 &  1 &  0 &  0 \\ 
 X_{12} &  0 & 1 &  0 &  0 &  0 &  1 &  0 &  0 &  0 &  0 &  1 &  0 &  1 &  0 &  1 \\ 
 X_{23} &  0 & 1 &  0 &  0 &  1 &  0 &  0 &  0 &  0 &  0 &  0 &  1 &  0 &  1 &  1 \\ 
 X_{34} &  0 & 1 &  0 &  0 &  0 &  1 &  0 &  0 &  0 &  1 &  0 &  0 &  1 &  1 &  0 \\ 
 X_{45} &  0 & 1 &  0 &  0 &  1 &  0 &  0 &  0 &  1 &  0 &  0 &  0 &  1 &  0 &  1 \\ 
 X_{56} &  0 & 1 &  0 &  0 &  0 &  1 &  0 &  1 &  0 &  0 &  0 &  0 &  0 &  1 &  1 \\ 
 X_{61} &  0 & 1 &  0 &  0 &  1 &  0 &  1 &  0 &  0 &  0 &  0 &  0 &  1 &  1 &  0 \\ 
 Y_{13} &  0 & 0 &  0 &  1  &  0 &  0 &  0 &  0 &  0 &  0 &  1 &  1 &  0 &  0 &  1 \\ 
 Y_{24} &  0 & 0 &  0 &  1  &  0 &  0 &  0 &  0 &  0 &  1 &  0 &  1 &  0 &  1 &  0 \\ 
 Y_{35} &  0 & 0 &  0 &  1 &  0 &  0 &  0 &  0 &  1 &  1 &  0 &  0 &  1 &  0 &  0 \\ 
 Y_{46} &  0 & 0 &  0 &  1 &  0 &  0 &  0 &  1 &  1 &  0 &  0 &  0 &  0 &  0 &  1 \\ 
 Y_{51} &  0 & 0 &  0 &  1 &  0 &  0 &  1 &  1 &  0 &  0 &  0 &  0 &  0 &  1 &  0 \\ 
 Y_{62} &  0 & 0 &  0 &  1 &  0 &  0 &  1 &  0 &  0 &  0 &  1 &  0 &  1 &  0 &  0 \\ 
 Z_{12} &  1 & 0 &  0 &  0 &  0 &  1 &  0 &  0 &  0 &  0 &  1 &  0 &  1 &  0 &  1 \\ 
 Z_{23} &  1 & 0 &  0 &  0 &  1 &  0 &  0 &  0 &  0 &  0 &  0 &  1 &  0 &  1 &  1 \\ 
 Z_{34} &  1 & 0 &  0 &  0  &  0 &  1 &  0 &  0 &  0 &  1 &  0 &  0 &  1 &  1 &  0 \\ 
 Z_{45} &  1 & 0 &  0 &  0 &  1 &  0 &  0 &  0 &  1 &  0 &  0 &  0 &  1 &  0 &  1 \\ 
 Z_{56} &  1 & 0 &  0 &  0 &  0 &  1 &  0 &  1 &  0 &  0 &  0 &  0 &  0 &  1 &  1 \\ 
 Z_{61} &  1 & 0 &  0 &  0  &  1 &  0 &  1 &  0 &  0 &  0 &  0 &  0 &  1 &  1 &  0
\ea
\right)
$}
,~
\eea
which in turn allows us to calculate the $Q_{JE}$- and $Q_D$-matrices using \eref{es03a04}.
The resulting toric diagram is encoded in the following matrix, 
\beal{es05a14}
G_t =
\resizebox{0.32\textwidth}{!}{$
\left(
\ba{cccc|cc|cccccc|ccc}
 p_1 & p_2 & p_3 & p_4 & q_1 & q_2 & s_1 & s_2 & s_3 & s_4 & s_5 & s_6 & o_1 & o_2 & o_3  \\
\hline
 -2 & 0 & 1 & 0 & -1 & -1 & 0 & 0 & 0 & 0 & 0 & 0 & -1 & -1 & -1 \\
 -2 & 0 & 0 & 1 & -1 & -1 & 0 & 0 & 0 & 0 & 0 & 0 & -1 & -1 & -1 \\
  -1 & 1 & 0 & 0 & 0 & 0 & 0 & 0 & 0 & 0 & 0 & 0 & 0 & 0 & 0 \\
 1 & 1 & 1 & 1 & 1 & 1 & 1 & 1 & 1 & 1 & 1 & 1 & 2 & 2 & 2 \\
\ea
\right)
$}
,~
\eea
where columns correspond to vertices in the toric diagram and the associated GLSM fields, as illustrated in \fref{fig_02}(a). 

The global symmetry of the brane brick model is $SU(2)_x \times SU(2)_y \times U(1)_f \times U(1)_R$, and the GLSM fields $p_1, p_2, p_3, p_4$ corresponding to the extremal vertices of the toric diagram in \fref{fig_02}(a) carry charges under the global symmetry as summarized in \tref{tab_10}.

\begin{table}[ht!]
\centering
\begin{tabular}{|c|c|c|c|c|l|}
\hline
\; & $SU(2)_x$ & $SU(2)_y$ & $U(1)_f$ & $U(1)_R$ & fugacity \\
\hline
$p_1$ & $+1$ & $0$ & $+1$ & $1/2$ & $t_1 = x f t$ \\
$p_2$ & $-1$ & $0$ & $+1$ & $1/2$ & $t_2=x^{-1} f t$\\
$p_3$ & $0$ & $+1$ & $-1$ & $1/2$ & $t_3=y f^{-1} t$\\
$p_4$ & $0$ & $-1$ & $-1$ &$1/2$ & $t_4=y^{-1} f^{-1} t$ \\
\hline
\end{tabular}
\caption{Global symmetries and charges of the $\mathbb{C}^{4}/\mathbb{Z}_6$ $(1,1,2,2)$ model. \label{tab_10}}
\end{table}

\begin{figure}
  \includegraphics[width=0.48\textwidth]{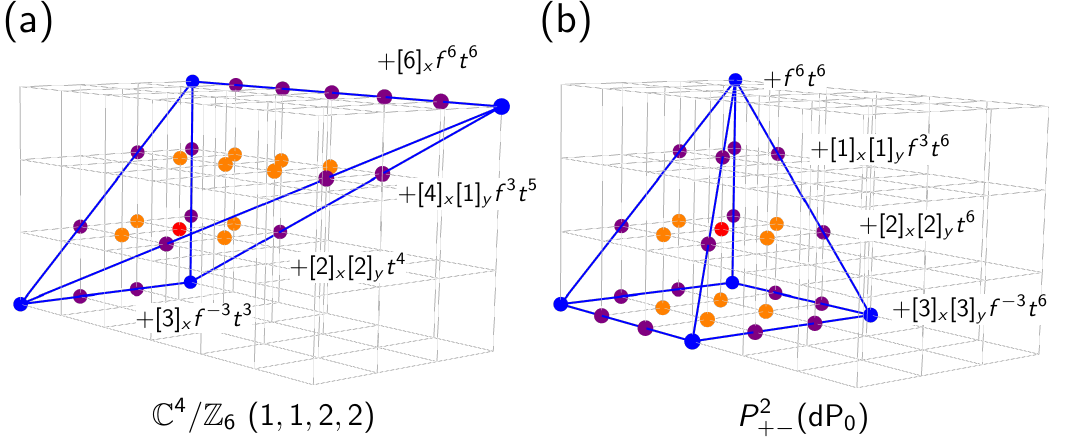}
  \caption{Lattice of generators for (a) the $\mathbb{C}^{4}/\mathbb{Z}_6$ $(1,1,2,2)$ model and (b) the $P_{+-}^{2}(\text{dP}_0)$ model, where $\mathbb{Z}^3$ layers are labelled by the corresponding positive terms in the respective plethystic logarithms of the Hilbert series. Both models have in total 30 generators, which form dual convex polytopes of the corresponding toric diagrams. \label{fig_03}}
\end{figure}

Using the Molien integral formula in \eref{es03a10}, the Hilbert series with global symmetry refinement for the mesonic moduli space of the $\mathbb{C}^{4}/\mathbb{Z}_6$ $(1,1,2,2)$ model takes the following form, 
\beal{es04a04}
&&
g(x,y,f,t;  \mathbf{C}^{4}/\mathbf{Z}_6)
=  
\nn\\
&&
\hspace{0.2cm}
\sum_{n_1, n_2=0}^{\infty}
[6n_1]_x [3n_2]_y f^{6n_1-3n_2 } t^{6n_1+3n_2 }
\nn\\
&&
\hspace{0.2cm}
+\sum_{n_1, n_2=0}^{\infty}
[6n_1+2]_x [3n_2+2]_y f^{6n_1-3n_2 } t^{6n_1+3n_2 +4}
\nn\\
&&
\hspace{0.2cm}
+\sum_{n_1, n_2=0}^{\infty}
[6n_1+4]_x [3n_2+1]_y f^{6n_1-3n_2 +3} t^{6n_1+3n_2 +5}
~,~
\nn\\
\eea
where $[m]_x [n]_y$ are characters of irreducible representations of $SU(2)_x \times SU(2)_y$ with highest weight $(m),(n)$, and $f$ and $t$ are the fugacities for $U(1)_f$ and $U(1)_R$, respectively. 
Here, the fugacity $t$ for $U(1)_R$ is scaled such that it corresponds to a $U(1)_R$ charge of $1/2$.

The corresponding plethystics logarithm of the Hilbert series takes the following form, 
\beal{es03a25}
&&
PL[g(x,y,f,t;  \mathbb{C}^{4}/\mathbb{Z}_6)]
=
[3]_x f^{-3} t^3
+ [2]_x [2]_y t^4
\nn\\
&&
\hspace{0.5cm}
+ [4]_x [1]_y f^3 t^5
+ [6]_x f^6 t^6
- [2]_y f^{-6} t^6
\nn\\
&&
\hspace{0.5cm}
- ([2]_x [2]_y f^{-3} + [2]_x [1]_y f^{-3})t^7
- ([4]_x [4]_y 
\nn\\
&&
\hspace{0.5cm}
+ [4]_x [2]_y + [2]_x [4]_y + [4]_x + [4]_y) t^8
+ \dots
~.~
\nn\\
\eea
The infinite expansion of the plethystic logarithm indicates that the mesonic moduli space here is a non-complete intersection.
There are in total 30 generators, whose mesonic flavor charges can be rescaled using the following fugacity map, 
\beal{es03a25b}
\tilde{x} = x^2~,~
\tilde{y} = y^2~,~
\tilde{f} = y^{-1} f^3~,~
\eea
such that the rescaled charges of the generators as summarized in \tref{tab_11} form a convex lattice polytope as illustrated in \fref{fig_03}(a).

\begin{table}[ht!]
\centering
\begin{tabular}{|c|c|ccc|}
\hline
PL term & generator & $SU(2)_{\tilde{x}}$ & $SU(2)_{\tilde{y}}$ & $U(1)_{\tilde{f}}$
\\
\hline 
\multirow{4}{*}{$+[3]_x f^{-3} t^3$} & $p_3^3 ~s o$  & $-1$ &$+1$ &$-1$\\
& $p_3^2 p_4 ~s o$  & $-1$ & $0$ & $-1$\\
& $p_3 p_4^2 ~s o$  & $-1$ & $-1$ & $-1$\\
& $p_4^3 ~s o$  & $-1$ & $-2$ & $-1$\\
\hline
\multirow{9}{*}{$+[2]_x [2]_y t^4$} & $p_1^2 p_3^2  ~q s o^2$  & $+1$ & $+1$ &$0$\\
&$p_1 p_2 p_3^2  ~q s o^2$  & $0$ &$+1$ & $0$\\
&$p_2^2 p_3^2  ~q s o^2$  & $-1$ & $+1$ & $0$\\
&$p_1^2 p_3 p_4  ~q s o^2$  & $+1$ & $0$ & $0$\\
&$p_1 p_2 p_3 p_4 ~q s o^2$  & $0$ & $0$ & $0$ \\
&$p_2^2 p_3 p_4  ~q s o^2$  & $-1$ & $0$ & $0$\\
&$p_1^2 p_4^2  ~q s o^2$  & $+1$ & $-1$ & $0$\\
&$p_1 p_2 p_4^2  ~q s o^2$  & $0$ & $-1$ & $0$ \\
&$p_2^2 p_4^2  ~q s o^2$  & $-1$ & $-1$ & $0$\\
\hline
\multirow{9}{*}{$+[4]_x [1]_y f^3 t^5$} &$p_1^4 p_3  ~q^2 s o^3$  & $+3$ & $+1$ & $+1$ \\
&$p_1^3 p_2 p_3  ~q^2 s o^3$  & $+2$ & $+1$ & $+1$\\
&$p_1^2 p_2^2 p_3  ~q^2 s o^3$  & $+1$ & $+1$ & $+1$\\
&$p_1 p_2^3 p_3  ~q^2 s o^3$  & $0$ & $+1$ & $+1$\\
&$p_2^4 p_3  ~q^2 s o^3$  & $-1$ & $+1$ & $+1$\\
&$p_1^4 p_4  ~q^2 s o^3$  & $+3$ & $0$ & $+1$\\
&$p_1^3 p_2 p_4  ~q^2 s o^3$  & $+2$ & $0$ & $+1$\\
&$p_1^2 p_2^2 p_4  ~q^2 s o^3$  & $+1$ & $0$ & $+1$\\
&$p_1 p_2^3 p_4  ~q^2 s o^3$  & $0$ & $0$ & $+1$\\
&$p_2^4 p_4  ~q^2 s o^3$  & $-1$ &$0$ & $+1$\\
\hline
\multirow{9}{*}{$+[6]_x f^6 t^6$} &$p_1^6  ~q^3 s o^4$  & $+5$ & $+1$ & $+2$ \\
&$p_1^5 p_2  ~q^3 s o^4$  & $+4$ & $+1$ & $+2$\\
&$p_1^4 p_2^2  ~q^3 s o^4$  & $+3$ & $+1$ & $+2$\\
&$p_1^3 p_2^3  ~q^3 s o^4$  & $+2$ & $+1$ & $+2$\\
&$p_1^2 p_2^4  ~q^3 s o^4$  & $+1$ & $+1$ & $+2$\\
&$p_1 p_2^5  ~q^3 s o^4$  & $0$ & $+1$ & $+2$\\
&$p_2^6  ~q^3 s o^4$  & $-1$ & $+1$ &h $+2$\\
\hline
\end{tabular}
\caption{Generators of the $\mathbb{C}^{4}/\mathbb{Z}_6$ $(1,1,2,2)$ model in terms of GLSM fields and their corresponding global flavor symmetry charges on the $\mathbb{Z}^3$ lattice. \label{tab_11}}
\end{table}

We can now deform the $\mathbb{C}^{4}/\mathbb{Z}_6$ $(1,1,2,2)$ model by introducing the following mass terms to the $J$- and $E$-terms in \eref{es05a10} as follows, 
\beal{es03a30}
(\Lambda_{15}, Y_{51}) ~:~
J_{51}^\prime &=&  + m Y_{51} + Z_{56} \cdot X_{61} - X_{56} \cdot Z_{61}
\nn\\
E_{15}^\prime &=& E_{15}
\nn\\
(\Lambda_{26}, Y_{62}) ~:~
J_{62}^\prime &=& - m Y_{62} + Z_{61} \cdot X_{12} - X_{61} \cdot Z_{12} 
\nn\\
E_{26}^\prime &=& E_{26} 
\nn\\
(\Lambda_{31}, Y_{13}) ~:~
J_{13}^\prime &=& + m Y_{13} + Z_{12} \cdot X_{23} - X_{12} \cdot Z_{23} 
\nn\\
E_{13}^\prime &=& E_{13}
\nn\\
(\Lambda_{42}, Y_{24}) ~:~
J_{24}^\prime &=& -m Y_{24} + Z_{23} \cdot X_{34} - X_{23} \cdot Z_{34} 
\nn\\
E_{42}^\prime &=& E_{42}
\nn\\
(\Lambda_{53}, Y_{35}) ~:~
J_{35}^\prime &=& +m Y_{35} + Z_{34} \cdot X_{45} - X_{34} \cdot Z_{45} 
\nn\\
E_{53}^\prime &=& E_{53}
\nn\\
(\Lambda_{64}, Y_{46}) ~:~
J_{46}^\prime &=& -m Y_{46} + Z_{45} \cdot X_{56} - X_{45} \cdot Z_{56} 
\nn\\
E_{64}^\prime &=& E_{64}
~.~
\eea
Here, we note that the massive chiral fields $\{Y_{51}, Y_{62}, Y_{13}, Y_{24}, Y_{35}, Y_{46}\}$ are all simultaneously part of the brick matching $p_4$ as shown in \eref{es05a11}.
We therefore refer to $p_4$ as the \textit{massive brick matching}, as first discussed in \cite{Franco:2023tyf}.
By integrating out the massive terms in the mass-deformed $J$- and $E$-terms, we obtain the $J$- and $E$-terms of the $P_{+-}^2(\text{dP}_0)$ model as follows, 
\beal{es03a40}
\begin{array}{rrcccc}
\Lambda^{1}_{14} :  & J^{1}_{41} & = &  Z_{45} \cdot Z_{56} \cdot X_{61} &-& X_{45} \cdot Z_{56} \cdot Z_{61} \\ 
\overline{\Lambda}^{1}_{14} :  & E^{1}_{14} & = &  D_{13} \cdot X_{34} &-& X_{12} \cdot D_{24} \\
\Lambda^{2}_{14} :  & J^{2}_{41} & = &  X_{45} \cdot X_{56} \cdot Z_{61} &-& Z_{45} \cdot X_{56} \cdot X_{61} \\  
\overline{\Lambda}^{2}_{14} :  & E^{2}_{14} & = &  D_{13} \cdot Z_{34} &-& Z_{12} \cdot D_{24} \\
\Lambda^{1}_{25} :  &  J^{1}_{52} &=& X_{56} \cdot Z_{61} \cdot Z_{12} &-& Z_{56} \cdot Z_{61} \cdot X_{12} \\ 
\overline{\Lambda}^{1}_{25} :  &  E^{1}_{25} &=&  D_{24} \cdot X_{45} &-& X_{23} \cdot D_{35} \\
\Lambda^{2}_{25} :  & J^{2}_{52} &= &  Z_{56} \cdot X_{61} \cdot X_{12} &-& X_{56} \cdot X_{61} \cdot Z_{12} \\ 
\overline{\Lambda}^{2}_{25} :  & E^{2}_{25} &= &  D_{24} \cdot Z_{45} &-& Z_{23} \cdot D_{35} \\
\Lambda^{1}_{36} :  & J^{1}_{63} &=&  Z_{61} \cdot Z_{12} \cdot X_{23} &-& X_{61} \cdot Z_{12} \cdot Z_{23} \\ 
\overline{\Lambda}^{1}_{36} :  & E^{1}_{36} &=&  D_{35} \cdot X_{56} &-& X_{34} \cdot D_{46} \\
\Lambda^{2}_{36} :  &  J^{2}_{63} &=& X_{61} \cdot X_{12} \cdot Z_{23} &-& Z_{61} \cdot X_{12} \cdot X_{23} \\
\overline{\Lambda}^{2}_{36} :  &  E^{2}_{36} &=&   D_{35} \cdot Z_{56} &-& Z_{34} \cdot D_{46} \\
\Lambda^{1}_{41} :  &  J^{1}_{14} &=& X_{12} \cdot Z_{23} \cdot Z_{34} &-& Z_{12} \cdot Z_{23} \cdot X_{34} \\ 
\overline{\Lambda}^{1}_{41} :  &  E^{1}_{41} &=&  D_{46} \cdot X_{61} &-& X_{45} \cdot D_{51} \\
\Lambda^{2}_{41} :  & J^{2}_{14} &=&  Z_{12} \cdot X_{23} \cdot X_{34} &-& X_{12} \cdot X_{23} \cdot Z_{34} \\ 
\overline{\Lambda}^{2}_{41} :  & E^{2}_{41} &=&  D_{46} \cdot Z_{61} &-& Z_{45} \cdot D_{51} \\
\Lambda^{1}_{52} :  &  J^{1}_{25} &=& Z_{23} \cdot Z_{34} \cdot X_{45} &-& X_{23} \cdot Z_{34} \cdot Z_{45} \\ 
\overline{\Lambda}^{1}_{52} :  &  E^{1}_{52} &=& D_{51} \cdot X_{12} &-& X_{56} \cdot D_{62} \\
\Lambda^{2}_{52} :  &  J^{2}_{25} &=& X_{23} \cdot X_{34} \cdot Z_{45} &-& Z_{23} \cdot X_{34} \cdot X_{45} \\
\overline{\Lambda}^{2}_{52} :  & E^{2}_{52} &=&  D_{51} \cdot Z_{12} &-& Z_{56} \cdot D_{62} \\
\Lambda^{1}_{63} :  & J^{1}_{36} &=&  X_{34} \cdot Z_{45} \cdot Z_{56} &-& Z_{34} \cdot Z_{45} \cdot X_{56} \\ 
\overline{\Lambda}^{1}_{63} :  & E^{1}_{63} &=&  D_{62} \cdot X_{23} &-& X_{61} \cdot D_{13} \\
\Lambda^{2}_{63} :  &  J^{2}_{26} &=& Z_{34} \cdot X_{45} \cdot X_{56} &-& X_{34} \cdot X_{45} \cdot Z_{56} \\ 
\overline{\Lambda}^{2}_{63} :  &  E^{2}_{63} &=&D_{62} \cdot Z_{23} &-& Z_{61} \cdot D_{13} \\
\end{array} ~.~
\nn\\
\eea
The corresponding quiver diagram after the mass deformation is shown in \fref{fig_045}(b). 
The new brick matching matrix is as follows,  
\beal{es03a41}
P=
\resizebox{0.35\textwidth}{!}{$
\left(
\ba{c|ccccc|cccccc|ccc}
\; &   p_2 & p_3 & p_4 & p_5 & p_6 & s_1 & s_2 & s_3 & s_4 & s_5 & s_6 & o_1 & o_2 & o_3  \\
\hline
 D_{13} &  0 &  1 &  0 &  0 &  0 &  0 &  0 &  0 &  0 &  1 &  1 &  0 &  0 &  1 \\ 
 D_{24} &  0 &  1 &  0 &  0 &  0 &  0 &  0 &  0 &  1 &  0 &  1 &  0 &  1 &  0 \\ 
 D_{35} &  0 &  1 &  0 &  0 &  0 &  0 &  0 &  1 &  1 &  0 &  0 &  1 &  0 &  0 \\ 
 D_{46} &  0 &  1 &  0 &  0 &  0 &  0 &  1 &  1 &  0 &  0 &  0 &  0 &  0 &  1 \\ 
 D_{51} &  0 &  1 &  0 &  0 &  0 &  1 &  1 &  0 &  0 &  0 &  0 &  0 &  1 &  0 \\ 
 D_{62} &  0 &  1 &  0 &  0 &  0 &  1 &  0 &  0 &  0 &  1 &  0 &  1 &  0 &  0 \\ 
 X_{12} &  1 &  0 &  0 &  0 &  0 &  0 &  0 &  0 &  0 &  1 &  0 &  1 &  0 &  1 \\ 
 X_{23} &  0 &  0 &  1 &  0 &  0 &  0 &  0 &  0 &  0 &  0 &  1 &  0 &  1 &  1 \\ 
 X_{34} &  1 &  0 &  0 &  0 &  0 &  0 &  0 &  0 &  1 &  0 &  0 &  1 &  1 &  0 \\ 
 X_{45} &  0 &  0 &  1 &  0 &  0 &  0 &  0 &  1 &  0 &  0 &  0 &  1 &  0 &  1 \\ 
 X_{56} &  1 &  0 &  0 &  0 &  0 &  0 &  1 &  0 &  0 &  0 &  0 &  0 &  1 &  1 \\ 
 X_{61} &  0 &  0 &  1 &  0 &  0 &  1 &  0 &  0 &  0 &  0 &  0 &  1 &  1 &  0 \\ 
 Z_{12} &  0 &  0 &  0 &  1 &  0 &  0 &  0 &  0 &  0 &  1 &  0 &  1 &  0 &  1 \\ 
 Z_{23} &  0 &  0 &  0 &  0 &  1 &  0 &  0 &  0 &  0 &  0 &  1 &  0 &  1 &  1 \\ 
 Z_{34} &  0 &  0 &  0 &  1 &  0 &  0 &  0 &  0 &  1 &  0 &  0 &  1 &  1 &  0 \\ 
 Z_{45} &  0 &  0 &  0 &  0 &  1 &  0 &  0 &  1 &  0 &  0 &  0 &  1 &  0 &  1 \\ 
 Z_{56} &  0 &  0 &  0 &  1 &  0 &  0 &  1 &  0 &  0 &  0 &  0 &  0 &  1 &  1 \\ 
 Z_{61} &  0 &  0 &  0 &  0 &  1 &  1 &  0 &  0 &  0 &  0 &  0 &  1 &  1 &  0
\ea
\right)
$}
~,~
\nn\\
\eea
which can be used to calculate the new $Q_{JE}$- and $Q_D$-matrices using \eref{es03a04}. 
The resulting toric diagram is given by the following $G_t$-matrix,
\beal{es03a42}
G_t=
\resizebox{0.35\textwidth}{!}{$
\left(
\ba{ccccc|cccccc|ccc}
 p_2 & p_3 & p_4 & p_5 & p_6 & s_1 & s_2 & s_3 & s_4 & s_5 & s_6 & o_1 & o_2 & o_3  \\
\hline
 0 & 1 & 0 & -1 & -1 & 0 & 0 & 0 & 0 & 0 & 0 & -1 & -1 & -1 \\
 0 & 0 & 1 & 0 & -1 & 0 & 0 & 0 & 0 & 0 & 0 & 0 & 0 & 0 \\
 1 & 0 & 0 & -1 & 0 & 0 & 0 & 0 & 0 & 0 & 0 & 0 & 0 & 0 \\
 1 & 1 & 1 & 1 & 1 & 1 & 1 & 1 & 1 & 1 & 1 & 3 & 3 & 3 \\
\ea
\right)
$}
~,~
\nn\\
\eea
where the columns correspond to vertices in the toric diagram associated to GLSM fields in the brane brick model. 
The toric diagram is illustrated in \fref{fig_02}(c) and we confirm that it corresponds to the $P_{+-}^2(\text{dP}_0)$ model.

\begin{table}[ht!]
\centering
\begin{tabular}{|c|c|c|c|c|l|}
\hline
\; & $SU(2)_x$ & $SU(2)_y$ & $U(1)_f$ & $U(1)_R$ & fugacity \\
\hline
$p_2$ & $+1$ & $0$ & $0$ & $1/3$ & $t_2=x t$\\
$p_3$ & $0$ & $0$ & $+2$ & $2/3$ & $t_3=f^{-2} t^2$ \\
$p_4$ & $0$ & $+1$ & $-1$ & $1/3$ &$t_4=yf^{-1} t$\\
$p_5$ & $-1$ & $0$ & $0$ &$1/3$&$t_5=x^{-1} t$ \\
$p_6$ & $0$ & $-1$ & $-1$ & $1/3$ & $t_6=y^{-1}f^{-1} t$\\
\hline
\end{tabular}
\caption{Global symmetries and charges of the $P_{+-}^{2}(\text{dP}_0)$ model after mass deformation. \label{tab_20}}
\end{table}

We can assign charges under the global symmetry $SU(2)_x \times SU(2)_y \times U(1)_f \times U(1)_R$ on the extremal brick matchings and the corresponding GLSM fields of the $P_{+-}^2(\text{dP}_0)$ model as summarized in \tref{tab_20}. 
Using the global charge assignment and the corresponding fugacities, we can calculate the Hilbert series of the mesonic moduli space as follows,
\beal{es03a45}
&&
g(x,y,f,t;  P_{+-}^{2}(\text{dP}_0))
=  
\nn\\
&&
\sum_{n_1, n_2 = 0}^{\infty}
[3n_2]_x [3n_2]_y f^{6n_1 - 3n_2} t^{6n_1 + 6n_2}
\nn\\
&&
+ \sum_{n_1, n_2 = 0}^{\infty}
[3n_2 + 1]_x [3n_2 +1]_y f^{6n_1 - 3n_2 + 3} t^{6n_1 + 6n_2 + 6}
\nn\\
&&
+ \sum_{n_1, n_2 = 0}^{\infty}
[3n_2 + 2]_x [3n_2 + 2]_y f^{6n_1 - 3 n_2} t^{6n_1 + 6n_2 + 6}
~,~
\nn\\
\eea
where as before, $[m]_x [n]_y$ are characters of irreducible representations of $SU(2)_x \times SU(2)_y$ with highest weight $(m),(n)$, and $f$ and $t$ are the fugacities for $U(1)_f$ and $U(1)_R$, respectively. 

The corresponding plethystic logarithm takes the following form,
\beal{es03a50}
&&
PL[g(x,y,b,t;  P_{+-}^{2}(\text{dP}_0))]
=
(f^6 
+ [1]_x [1]_y f^3 
\nn\\
&&
\hspace{0.5cm}
+ [2]_x [2]_y 
+ [3]_x [3]_y f^{-3}  )t^6
- (f^6 + [2]_x [2]_y f^6
\nn\\
&&
\hspace{0.5cm}
+ [3]_x [3]_y f^3 + [1]_x [3]_y f^3 + [3]_x [1]_y f^3 )t^{12}
+ \dots
~,~
\nn\\
\eea
where we note that the expansion is infinite and hence the mesonic moduli space is a non-complete intersection. 
As in the case of the $\mathbb{C}^{4}/\mathbb{Z}_6$ $(1,1,2,2)$ model, the number of generators for the $P_{+-}^2(\text{dP}_0)$ model after mass deformation remains 30, indicating that the number of generators is an invariant under a mass deformation that corresponds to an algebraic and combinatorial polytope mutation of a toric Fano 3-fold. 
The mesonic flavor symmetry charges on the generators can be rescaled using the following fugacity map, 
\beal{es03a55}
\tilde{x}= x^2~,~
\tilde{y}= y^2~,~
\tilde{f}= x^{-1} y^{-1} f^3~,~
\eea
such that the mesonic flavor charges of the generators form a convex lattice polytope in $\mathbb{Z}^3$ as illustrated in \fref{fig_03}(b).
We note here that the lattice of generators forms a reflexive polytope which is dual to the toric diagram of $P_{+-}^2(\text{dP}_0)$ in \fref{fig_02}(c). 
\tref{tab_21} summarizes the generators in terms of GLSM fields with the corresponding rescaled mesonic flavor charges. 

\begin{table}[ht!]
\centering
\begin{tabular}{|c|c|ccc|}
\hline
PL term & generator & $SU(2)_{\tilde{x}}$ & $SU(2)_{\tilde{y}}$ & $U(1)_{\tilde{f}}$
\\
\hline 
$+ f^6 t^6$
&$p_3^3~ s o$  & $+1$ & $+1$ & $+2$\\
\hline
\multirow{4}{*}{$+[1]_x [1]_y f^3 t^6$} 
&$p_2 p_3^2 p_4~ s o^2$  & $+1$ & $+1$ & $+1$\\
&$p_3^2 p_4 p_5~ s o^2$  & $0$ & $+1$ & $+1$\\
&$p_2 p_3^2 p_6~ s o^2$  & $+1$ & $0$ & $+1$\\
&$p_3^2 p_5 p_6~ s o^2$  & $0$ & $0$ & $+1$\\
\hline
\multirow{9}{*}{$+[2]_x [2]_y t^6$} 
&$p_2^2 p_3 p_4^2~ s o^3$  & $+1$ & $+1$ & $0$\\
&$p_2 p_3 p_4^2 p_5~ s o^3$  & $0$ & $+1$ & $0$\\
&$p_3 p_4^2 p_5^2~ s o^3$  & $-1$ & $+1$ & $0$\\
&$p_2^2 p_3 p_4 p_6~ s o^3$  & $+1$ & $0$ & $0$\\
&$p_2 p_3 p_4 p_5 p_6~ s o^3$  & $0$ & $0$ & $0$ \\
&$p_3 p_4 p_5^2 p_6~ s o^3$  & $-1$ & $0$ & $0$ \\
&$p_2^2 p_3 p_6^2~ s o^3$  & $+1$ & $-1$ & $0$\\
&$p_2 p_3 p_5 p_6^2~ s o^3$  & $0$ & $-1$ & $0$\\
&$p_3 p_5^2 p_6^2~ s o^3$  & $-1$ & $-1$ & $0$\\
\hline
\multirow{16}{*}{$+[3]_x [3]_y f^{-3} t^6$} 
&$p_2^3 p_4^3~ s o^3$  & $+1$ & $+1$ & $-1$  \\
&$p_2^2 p_4^3 p_5~ s o^3$ & $0$ & $+1$ & $-1$\\
&$p_2 p_4^3 p_5^2~ s o^3$  & $-1$ & $+1$ & $-1$\\
&$p_4^3 p_5^3~ s o^3$  & $-2$ & $+1$ & $-1$\\
&$p_2^3 p_4^2 p_6~ s o^3$  & $+1$ & $0$ & $-1$\\
&$p_2^2 p_4^2 p_5 p_6~ s o^3$  & $0$ & $0$ & $-1$\\
&$p_2 p_4^2 p_5^2 p_6~ s o^3$  & $-1$ & $0$ & $-1$\\
&$p_4^2 p_5^3 p_6~ s o^3$  & $-2$ & $0$ & $-1$\\
&$p_2^3 p_4 p_6^2~ s o^3$  & $+1$ & $-1$ & $-1$\\
&$p_2^2 p_4 p_5 p_6^2~ s o^3$  & $0$ & $-1$ & $-1$\\
&$p_2 p_4 p_5^2 p_6^2~ s o^3$  & $-1$ & $-1$ & $-1$\\
&$p_4 p_5^3 p_6^2 ~ s o^3$  & $-2$ & $-1$ & $-1$\\
&$p_2^3 p_6^3 ~ s o^3$  & $+1$ & $-2$ & $-1$\\
&$p_2^2 p_5 p_6^3 ~ s o^3$  & $0$ & $-2$ & $-1$\\
&$p_2 p_5^2 p_6^3 ~ s o^3$  & $-1$ & $-2$ & $-1$\\
&$p_5^3 p_6^3 ~ s o^3$  & $-2$ & $-2$ & $-1$\\
\hline
\end{tabular}
\caption{Generators of the $P_{+-}^{2}(\text{dP}_0)$ model after mass deformation in terms of GLSM fields and their corresponding global flavor symmetry charges on the $\mathbb{Z}^3$ lattice. \label{tab_21}}
\end{table}

We note here that the generators of the mesonic moduli space for the $P_{+-}^2(\text{dP}_0)$ model have all the same overall $U(1)_R$ charge, as indicated by the common exponent of the $U(1)_R$ fugacity $t$ for the first positive terms in the plethystic logarithm in \eref{es03a45}. 
This implies that the lattice of generators illustrated in \fref{fig_03}(b) is a 3-dimensional plane embedded in a 4-dimensional space of charges under the full global symmetry $SU(2)_x \times SU(2)_y \times U(1)_f \times U(1)_R$.
We can also ensure in the $\mathbb{C}^{4}/\mathbb{Z}_6$ $(1,1,2,2)$ model that the generators of the mesonic moduli space have the same $U(1)_R$ charge by imposing the $U(1)_R$ charges on the extremal brick matchings $p_2, \dots, p_6$ of the $P_{+-}^2(\text{dP}_0)$ model in \tref{tab_10} on the extremal brick matchings $p_1, \dots, p_4$ of the $P_{+-}^2(\text{dP}_0)$ model.
Here, $p_1$ of $P_{+-}^2(\text{dP}_0)$ gets an $U(1)_R$ charge of $2/3$ such that the mass terms in \eref{es03a30} have all the same $U(1)_R$ charge assignment under the $U(1)_R$ charges from the $P_{+-}^2(\text{dP}_0)$ model. 
Under this $U(1)_R$ charge assignment, both the $\mathbb{C}^{4}/\mathbb{Z}_6$ $(1,1,2,2)$ model and the $P_{+-}^2(\text{dP}_0)$ model end up having the \textit{same} unrefined Hilbert series of the following form, 
\beal{es03a60}
g(t) = 
\frac{1 + 26 t^6 + 26 t^{12} + t^{18}
}{
(1-t^6)^4
}
~,~
\eea
where $t$ here is the remaining $U(1)_R$ fugacity, scaled to correspond to an $U(1)_R$ charge of $1/3$.
This unrefined Hilbert series common to the brane brick models before and after mass deformation has been referred to as the \textit{Ehrhart series} in \cite{akhtar2012minkowski}.
\\

\section{Discussions and Conclusions}\label{s_conclusion}

In this work, we observe that algebraic and combinatorial polytope mutation for toric Fano 3-folds is related to mass deformation of corresponding $2d$ $(0,2)$ supersymmetric gauge theories realized by brane brick models.
This correspondence appears to exist when particular conditions are met with various physically and mathematically interesting properties. 
Given that two abelian brane brick models are related by a mass deformation and correspond to toric Fano 3-folds related by a combinatorial and algebraic polytope mutation, we can observe the following conditions and properties:
\begin{itemize}
\item Let us assume that the combinatorial polytope mutation is defined by $\mu_{\textbf{w}}(\Delta, F)$ acting on the reflexive toric diagram $\Delta$ of the original toric Fano 3-fold.
The combinatorial polytope mutation $\mu_{\textbf{w}}(\Delta, F)$ and the corresponding algebraic polytope mutation $\varphi_A$
can be associated to a mass deformation if the original toric diagram $\Delta$ has $\text{width}_{\textbf{w}}(\Delta) = 2$ along the mutation direction $\textbf{w}$.

\item Assuming that the single internal point of the original reflexive toric diagram $\Delta$ is at the origin $(0,0,0)$ with height $h=0$ along the mutation vector $\textbf{w}$, massive brick matchings correspond to vertices in $\Delta$ with height $h=+1$ along $\mathbf{w}$, whereas brick matchings that move under mass deformation have height $h=-1$ along $\mathbf{w}$.
As discussed in \cite{Franco:2023tyf}, massive brick matchings contain all chiral fields in the brane brick model that become massive during mass deformation.

\item The two abelian brane brick models related by mass deformation have mesonic moduli spaces with the same number of generators. 
The lattice of generators formed by the mesonic flavor charges of the generators are dual reflexive polytopes of the toric diagrams of the associated Fano 3-folds. 

\item The Hilbert series of the mesonic moduli spaces of brane brick models related by mass deformation and algebraic and combinatorial polytope mutation are identical when unrefined. Here, unrefined means that both Hilbert series are in terms of the $U(1)_R$ symmetry fugacity $t$, where for the brane brick model prior to the mass deformation the $U(1)_R$ charge assignment on chiral fields is based on $U(1)_R$ charge assignment in the brane brick model after the mass deformation. 
\end{itemize}
We expect the above observations to hold generally for a large family of toric Fano 3-folds and corresponding brane brick models that exhibit algebraic and combinatorial polytope mutation with an associated mass deformation. 
We would like to emphasize that we are working on an extensive collection of other examples of mass deformations between brane brick models that correspond to algebraic and combinatorial polytope mutation and we hope to report on these cases in the near future. 

As a final comment, our results here are in line with recent developments in the context of mutations appearing for toric Calabi-Yau 3-folds, and their connection to generalized toric polygons (GTPs) \cite{Benini:2009gi}, $5d$ superconformal field theories and Hanany-Witten transitions \cite{Arias-Tamargo:2024fjt,Bourget:2023wlb,Franco:2023flw,Franco:2023mkw,Cremonesi:2023psg,higashitani2022deformations}. 
In the context of brane brick models and toric Fano 3-folds, our observations in this work
that $\text{width}_{\textbf{w}}(\Delta) = 2$ toric diagrams lead to a correspondence between mass deformation and polytope mutation, that the number of generators of the mesonic moduli space remains invariant under such deformations and mutations, and that the unrefined Hilbert series remains invariant before and after such deformations and mutations when the $U(1)_R$ charges after mass deformation are imposed, very much also apply to $4d$ $\mathcal{N}=1$ supersymmetric gauge theories associated to toric Calabi-Yau 3-folds and realized by brane tilings \cite{Franco:2005rj,Hanany:2005ve,Franco:2005sm,Hanany:2012hi}.
We plan to further investigate these cross-dimensional connections in the near future.
\\

\acknowledgments

D. G. would like to thank UNIST where this project was initiated.
He is supported by JST PRESTO Grant Number JPMJPR2117.
R.-K. S. would like to thank IBS Center for Geometry and Physics in Pohang, the Yau Mathematical Sciences Center at Tsinghua University in Beijing, as well as the Kavli Institute for the Physics and Mathematics of the Universe in Tokyo for their hospitality during various stages of this work. 
He is supported by a Basic Research Grant of the National Research Foundation of Korea (NRF-2022R1F1A1073128).
He is also supported by a Start-up Research Grant for new faculty at UNIST (1.210139.01) and a UNIST AI Incubator Grant (1.240022.01).  
He is also partly supported by the BK21 Program (``Next Generation Education Program for Mathematical Sciences'', 4299990414089) funded by the Ministry of Education in Korea and the National Research Foundation of Korea (NRF).


\bibliographystyle{jhep}
\bibliography{mybib}

\end{document}